\documentclass[structabstract]{aa}

\usepackage{graphicx}
\usepackage{txfonts}

\usepackage{natbib}

\newcommand{\Jupiter}{{\rm J}}

\newcommand{\Sun}{\odot}
\newcommand{\ddeg}{\hbox{.\hskip-2pt $^\circ$}}

\begin{document}

\title{The barycentric motion of exoplanet host stars}
\subtitle{Tests of solar spin--orbit coupling}

   \author{M.A.C. Perryman
          \inst{1,2}
          \and
          T. Schulze-Hartung\inst{2}
          }

   \institute{Astronomisches Rechen-Institut, Zentrum f\"ur Astronomie der Universit\"at Heidelberg, M\"onchhofstr.\ 12--14, D-69120 Heidelberg\\
         \and
             Max-Planck-Institut f\"ur Astronomie, K\"onigstuhl 17, D-69117 Heidelberg \\
             }

   \date{Received 1 September 2010; accepted 1 October 2010}

\abstract
{Empirical evidence suggests a tantalising but unproven link between various indicators of solar activity and the barycentric motion of the Sun. The latter is exemplified by transitions between regular and more disordered motion modulated by the motions of the giant planets, and rare periods of retrograde motion with negative orbital angular momentum. An examination of the barycentric motion of exoplanet host stars, and their stellar activity cycles, has the potential of proving or disproving the Sun's motion as an underlying factor in the complex patterns of short- and long-term solar variability indices, by establishing whether such correlations exist in other planetary systems. In either case, these studies may lead to further insight into the nature of the solar dynamo.}
{Some 40 multiple exoplanet systems are now known, all with reasonably accurate orbital elements. The forms and dynamical functions of the barycentric motion of their host stars are examined. These results can be compared with long-term activity indicators of exoplanet host stars, as they become available, to examine whether the correlations claimed for the Sun also exist in other systems.}
{Published orbital elements of multiple exoplanetary systems are used to examine their host star barycentric motions. For each system, we determine analytically the orbital angular momentum of the host star, and its rate of change.}
{A variety of complex patterns of barycentric motions of exoplanet host stars is demonstrated, depending on the number, masses and orbits of the planets. Each of the behavioural types proposed to correlate with solar activity are also evident in exoplanet host stars: repetitive patterns influenced by massive multiple planets, epochs of rapid change in orbital angular momentum, and intervals of negative orbital angular momentum.}
{The study provides the basis for independent investigations of the widely-studied but unproven suggestion that the Sun's motion is somehow linked to various indicators of solar activity. We show that, because of the nature of their barycentric motions, the host stars \object{HD~168443} and \object{HD~74156} offer particularly powerful tests of this hypothesis.}

\keywords{astrometry -- planets and satellites : general -- Sun : activity -- Stars : activity}

\maketitle
%________________________________________________________________

%--------------------
\section{Introduction}
%--------------------

The centre of mass of the solar system, the barycentre, moves through space controlled by the Galaxy's gravitational potential. For solar system dynamical studies, the barycentre is the origin of a local quasi-inertial frame. The Sun, the planets, and all other gravitating matter orbit this centre of mass, resulting in a continuous displacement of the Sun's position relative to the barycentre. This displacement is determinable, with adequate precision for the objectives of this study, by Newtonian dynamics.

As noted in the first discussions of the Sun's orbital motion about the barycentre by Newton \citep[as quoted by][]{1934cajori}, the motion of the Sun is rather complex {\it `since that centre of gravity is continually at rest, the Sun, according to the various positions of the planets, must continuously move every way, but will never recede far from that centre.'}

\citet{1965AJ.....70..193J} used the improved numerical integration of the outer solar system planets over about 400~years by \citet{1951USNAO..12....1E} to determine the path of the Sun with respect to the solar system barycentre more precisely than had been possible previously. He claimed a 178.7\,yr periodicity in its principal features, and suggested correlations of this, and the rate of change of the Sun's orbital angular momentum, with both the 11-year sunspot cycle, and the 22-year magnetic dipole inversion \citep{1924hale} and sunspot polarity cycles. \citet{2000SoPh..191..201J} criticised the latter claim on the basis of Jose's arbitrary phase adjustments. Whether these specific correlations exist or not, by drawing attention to various periodicities in the Sun's barycentric motion \citet{1965AJ.....70..193J} nevertheless stimulated further research into its possible link with solar activity. 

A substantial body of work has since appeared on the purported link between the Sun's motion and aspects of its long-term activity. With only one exemplar, the subject has reached something of an impasse. We show that exoplanet systems provide numerous other examples to allow further progress.

%--------------------
\section{The case of the solar system}
%--------------------

In the simplest case of a single orbiting planet, the star and planet displacements from the system barycentre are related by
\begin{equation}
{\bf r}_\star(t)= -\left(\frac{M_{\rm p}}{M_\star}\right)\; {\bf r}_{\rm p}(t) \ ,
\end{equation}
where $M_{\rm p}$ and $M_\star$ are the planet and star masses in common units, and ${\bf r}_{\rm p}$ and ${\bf r}_\star$ are the positions of the planet and star with respect to the barycentre, again in common units. Both planet and star move in elliptical orbits in the same plane, about the common centre of mass, with semi-major axes which differ by the ratio $M_{\rm p}/M_\star$, and with orientations (specified by their argument of pericentre, $\omega$) which differ by $\pi$~rad.

In a multiple planetary system in which the planets are gravitationally non-interacting, and in which the individual and total planet masses are small with respect to the mass of the star, the motion of the star about the system barycentre can be approximated by the linear superposition of the reflex motions due to the Keplerian orbit of each individual planet around that star--planet barycentre. If the planets have periods or close approaches such that they are dynamically interacting, the orbits must then be determined, at least over the longer term, by N-body integration. This is a complication that we ignore in this study.

Equation~1 shows that low-mass, short-period planets naturally have a smaller effect on the Sun's (or star's) displacement compared with more massive planets, and particularly those with large semi-major axes (i.e.\ long orbital periods). 

In the case of the solar system, the Sun's motion under the influence of the four giant planets (Jupiter, Saturn, Uranus, and Neptune) is illustrated in Figure~\ref{fig:orbits}a for the time interval 1970--2030. Equivalent plots, with which our own determinations are consistent, can be found in, e.g., \citet[][Figure~1]{1965AJ.....70..193J} over the interval 1834--2013, and in \citet[][Figure~2]{2000SoPh..191..201J} over the intervals 1815--65 and 1910--60. 

The Sun's barycentric motion is evidently most strongly influenced by the motions of these four giant planets. Patterns in rotationally invariant quantities, such as successive close approaches between the Sun and the system barycentre, termed `peribacs' by \citet{1987SoPh..110..191F}, recur, for example, at mean intervals of 19.86\,yr, corresponding to the synodic periodic of Jupiter and Saturn.

%--------------------
\subsection{Relation to solar activity}
%--------------------

Solar axial rotation plays a fundamental role in the two main hypotheses for a mechanism underlying the solar cycle: attributed to a turbulent dynamo operating in or below the convection envelope, or to a large-scale oscillation superimposed on a fossil magnetic field in the radiative core. The precise nature of the dynamo, and many of the details of the associated solar activity (such as the details of the sunspot cycles, or the reasons for the prolonged Maunder-type solar minima) remain unexplained \citep[e.g.][]{2005LRSP....2....2C, 2007AIPC..919...49C, 2009SSRv..144...53W, 2010SSRv..152..591J}, although certain features may arise naturally in some models \citep[e.g.][]{2007ApJ...658..657C}.

Empirical investigations have long pointed to a link between the Sun's barycentric motion and various solar variability indices \citep[e.g.][]{1859MNRAS..19...85W, 1900MNRAS..60..599B, 1911RSPSA..85..309S, 1969JBAA...79..385F}. 

For the reasons noted below, such a connection has not attracted much serious attention in theories developed to understand the solar dynamo. Nevertheless, acceleration in the Sun's motion, or the change of its orbital angular momentum, have been linked empirically to 
phenomena such as the Wolf sunspot number counts \citep{1965Natur.208..129W}, 
climatic changes \citep{1979stiw.conf..193M},
the 80--90-yr secular Gleissberg cycles \citep{1981JICR...12....3L, 1999SoPh..189..413L},
the prolonged Maunder-type solar minima \citep{1987SoPh..110..191F,1990BAICz..41..200C, 2000AnGeo..18..399C}, 
short-term variations in solar luminosity \citep{1990SoPh..127..379S}, 
sunspot extrema \citep{1999SoPh..189..413L}, 
the 2400-yr cycle seen in $^{14}$C tree-ring proxies \citep{2000AnGeo..18..399C}, 
hemispheric sunspot asymmetry \citep{2000SoPh..191..201J}, 
torsional oscillations in long-term sunspot clustering \citep{2003A&A...399..731J}, 
and violations of the Gnevishev--Ohl sunspot rule \citep{2005MNRAS.362.1311J}. 

We expand briefly on two of these themes to set the context. One of the specific curiosities of the Sun's motion noted by \citet{1965AJ.....70..193J} is evident in Figure~\ref{fig:orbits}a. Around 1990, and before that in 1811 and 1632, the Sun had a retrograde motion relative to the barycentre, i.e.\ its angular momentum with respect to the centre of mass was negative \citep{1965AJ.....70..193J}. The next such retrograde Sun event will occur around 2169. In a multi-body system in which total angular momentum is conserved, such behaviour of one particular body is, again, not in itself surprising. The result is of more particular interest because \citet{2005MNRAS.362.1311J} has argued that epochs violating the Gnevishev--Ohl `sunspot rule', which states that the sum of sunspot numbers over an odd-numbered cycle exceeds that of its preceding even-numbered cycle \citep{1948gnevishev, 2001ApJ...554L.119K, 2007ApJ...658..657C, 2009AstL...35..564N}, are close to these intervals of the Sun's retrograde orbital motion. 

% different authors give different dates, especially for Sporer
\citet{2000AnGeo..18..399C} drew attention to periods of the Sun's orbit which are characterised by more ordered `trefoil'-type motion dominated by Jupiter and Saturn, and which recur with the periodicity of 178.7~years. More `disordered' motion appears between these periods (their Figure~2). \citet{2000AnGeo..18..399C} associated these intervals with the prolonged minima of the solar activity cycle, viz.\ the Wolf (1280--1350), Sp\"orer (1430--1520), Maunder (1645--1715), and Dalton (1790--1820) minima.

%-----------------------------------
\begin{figure}[t]
\centering
\includegraphics[width=0.78\linewidth]{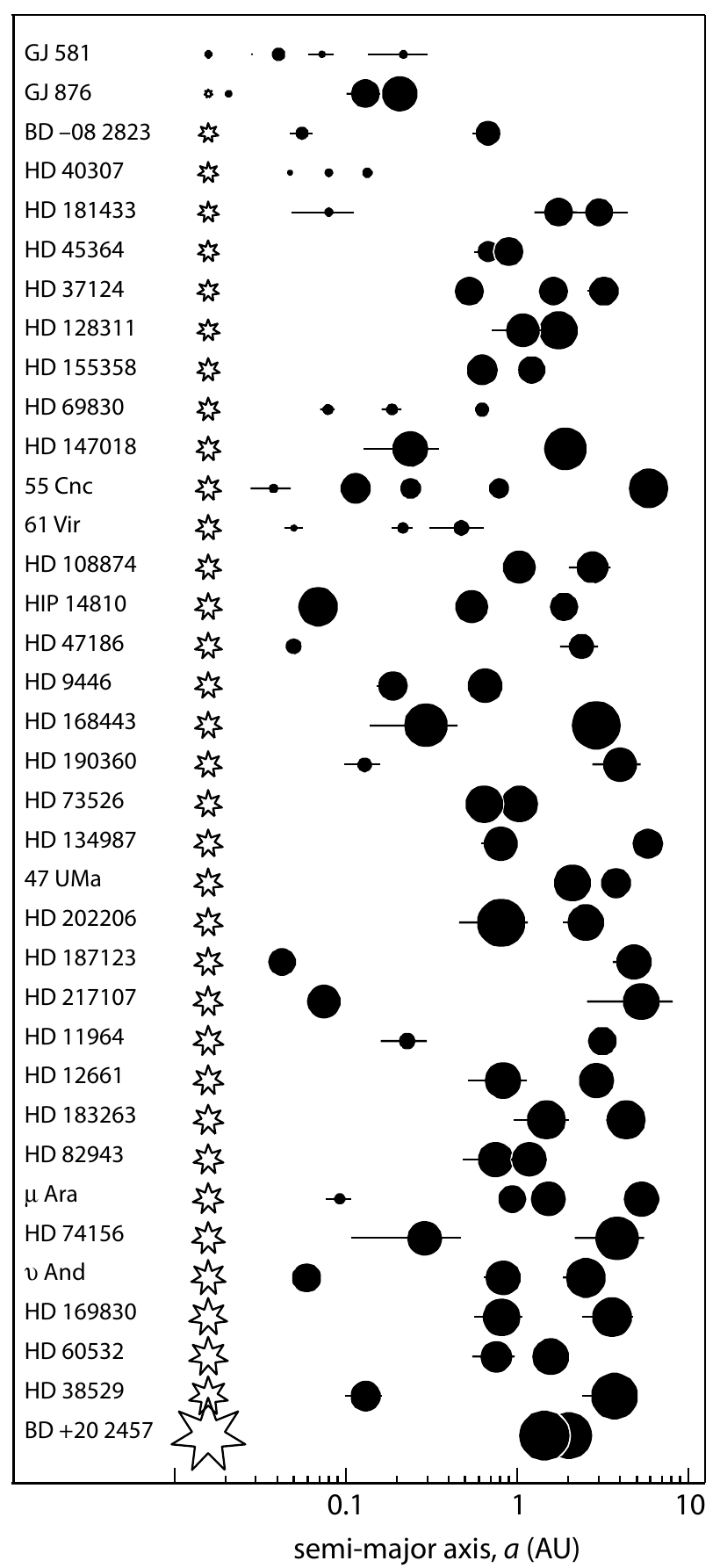}
\caption{Multiple (Doppler) planet systems, ordered by host star mass (indicated at left with size proportional to $M_\star$, ranging from $0.31M_\Sun$ for GJ~581 to $2.8M_\Sun$ for BD\,+20\,2457). Each planet in the system is shown to the right, with sizes proportional to $\log M_{\rm p}$ (ranging from about $0.01-20M_\Jupiter$). Horizontal bars through the planets indicate maximum and minimum star--planet distance based on their eccentricities. 
\label{fig:multiples}
}
\end{figure}
%-----------------------------------

%-----------------------------------
\begin{table}[t]
\caption{Barycentric host star motions, ordered by increasing $\vert\,{\rm d}L_z/{\rm d}t\,\vert_{\rm max}$. Maximum and minimum values of the star's orbital angular momentum are given for the time interval 2000--2100. $1\,M_\Sun\,{\rm AU}^2\,{\rm d}^{-1}\equiv5.15\times10^{47}$\,kg\,m$^2$\,s$^{-1}$; $1\,M_\Sun\,{\rm AU}^2\,{\rm d}^{-2}\equiv5.96\times10^{42}$\,kg\,m$^2$\,s$^{-2}$. 
}
\label{tab:systems} 
\centering
\begin{tabular}{lccc}
\hline\hline
\noalign{\vspace{2pt}}
System& 			$L_{z, \rm min}$&  				$L_{z, \rm max}$&  				$\vert\,{\rm d}L_z/{\rm d}t\,\vert_{\rm max}$  \\  
&				($M_\Sun\,{\rm AU}^2\,{\rm d}^{-1}$)&	$(M_\Sun\,{\rm AU}^2\,{\rm d}^{-1}$)&	$(M_\Sun\,{\rm AU}^2\,{\rm d}^{-2}$) \\
\hline
\noalign{\vspace{2pt}}
HD 40307		&	$-$1.51E--12	&	2.36E--11	&	9.33E--12	\\
CoRoT--7		&	$+$2.91E--13 	&	6.58E--12	&	1.78E--11	\\
Sun			&	$+$2.33E--09 	&	7.86E--08	&	3.91E--11	\\
HD 45364		&	$+$3.66E--09 	&	1.17E--08	&	7.67E--11	\\
HD 69830		&	$-$7.05E--11	&	2.14E--10	&	8.00E--11	\\
GJ 581		&	$-$3.17E--11	&	1.06E--10	&	8.06E--11	\\
61 Vir		&	$-$4.65E--11	&	1.90E--10	&	1.13E--10	\\
47 UMa		&	$+$7.21E--08 	&	2.29E--07	&	3.00E--10	\\
HD 155358	&	$-$1.93E--09	&	3.33E--08	&	5.22E--10	\\
HD 108874	&	$-$2.27E--08	&	1.44E--07	&	1.39E--09	\\
BD --08 2823	&	$+$6.06E--10 	&	2.61E--09	&	1.46E--09	\\
HD 11964		&	$+$3.42E--09 	&	1.66E--08	&	1.67E--09	\\
HD 215497	&	$+$6.39E--10 	&	2.72E--09	&	2.20E--09	\\
HD 37124		&	$-$2.52E--08	&	1.20E--07	&	2.46E--09	\\
HD 60532		&	$+$1.78E--08 	&	2.00E--07	&	3.38E--09	\\
HD 128311	&	$-$1.15E--10	&	4.84E--07	&	4.80E--09	\\
HD 82943		&	$-$1.52E--08	&	2.34E--07	&	5.09E--09	\\
HD 134987	&	$-$9.03E--08	&	2.36E--07	&	5.42E--09	\\
HD 73526		&	$-$1.62E--08	&	4.28E--07	&	6.48E--09	\\
HD 47186		&	$-$2.52E--09	&	7.94E--09	&	8.44E--09	\\
HD 181433	&	$-$1.30E--08	&	4.62E--08	&	9.27E--09	\\
HD 9446		&	$+$5.31E--09 	&	8.62E--08	&	1.02E--08	\\
GJ 876		&	$+$1.81E--08 	&	8.32E--08	&	1.05E--08	\\
HD 190360	&	$+$4.41E--08 	&	8.71E--08	&	1.10E--08	\\
HD 12661		&	$-$1.67E--07	&	5.07E--07	&	1.34E--08	\\
HD 183263	&	$-$4.16E--07	&	1.69E--06	&	1.90E--08	\\
$\mu$~Ara	&	$-$1.72E--07	&	5.76E--07	&	2.37E--08	\\
HD 169830	&	$-$4.38E--07	&	1.51E--06	&	4.76E--08	\\
BD +20 2457	&	$-$1.08E--07	&	1.59E--05	&	1.08E--07	\\
HD 202206	&	$+$6.70E--07 	&	6.38E--06	&	1.57E--07	\\
HD 147018	&	$-$4.50E--07	&	2.58E--06	&	5.05E--07	\\
HIP 14810	&	$-$3.29E--07	&	6.72E--07	&	5.82E--07	\\
55 Cnc		&	$-$6.43E--07	&	1.84E--06	&	6.29E--07	\\
$\upsilon$~And&	$-$5.43E--07	&	1.33E--06	&	7.80E--07	\\
HD 187123	&	$-$3.21E--07	&	5.77E--07	&	9.23E--07	\\
HD 38529		&	$+$1.74E--06 	&	6.06E--06	&	1.52E--06	\\
HD 217107	&	$-$1.21E--06	&	2.03E--06	&	1.58E--06	\\
HD 74156		&	$-$1.73E--06	&	4.70E--06	&	1.82E--06	\\
HAT--P--13	&	$+$8.57E--07 	&	4.48E--06	&	3.91E--06	\\
HD 168443	&	$-$1.07E--05	&	2.99E--05	&	6.42E--06	\\
\hline
\end{tabular}
\end{table}
%-----------------------------------

%--------------------
\subsection{The question of a causal link}
%--------------------

The implication of any relation between solar motion and solar activity, should it exist, is that there is some mechanism related to its motion which, in some way, alters or modulates material flows and/or magnetic fields within the rotating and revolving Sun. In Newtonian gravity, only planetary tidal forces ($\propto r^{-3}$) act externally to stress or distort the figure of the Sun, and these have been excluded as the underlying causal factor because of their small amplitudes \citep{1975Natur.253..511O, 1977Natur.266..434S}. In metric theories, that the Sun is in gravitational free fall has also been used to argue that there should be no causal link between its barycentric motion and any type of solar activity indicator \citep{2006MNRAS.368..280S}. 

Attempts to identify a spin--orbit coupling mechanism between the solar axial rotation and orbital revolution nevertheless continue. Thus \citet{1997ApJ...487..930Z} has suggested that the Sun's predominantly elliptical motion creates a periodic velocity shear on the weak poloidal magnetic field in its interior. \citet{2000SoPh..191..201J} has instead proposed that the orientation of the Sun's rotation axis (inclined by $\sim$7\ddeg25 with respect to the ecliptic, and by $\sim$6\ddeg25 with respect to the invariant plane, viz.\ the plane defined by the total angular momentum of the planets and the Sun) facilitates non-linear frequency mixing of planetary-induced forcing oscillations, and induces differential flows to conserve angular momentum. 

Although elements of each explanation have been criticised by \citet{2006MNRAS.368..280S}, the latter author nevertheless argues that a clear physical picture, properly linking the Sun's rotation and its revolution about the barycentre, remains elusive. 

We can summarise by stating that an empirical connection between solar activity and the Sun's motion appears intriguing, but remains unproven and unexplained. Perhaps like the Titius--Bode law, which has been largely attributed to phenomenological patterns which can be replicated by broad families of two-parameter geometrical progressions \citep[][Section~1.5]{1998Icar..135..549H, 2000ssd..book.....M}, it is purely specious. But it has been extensively hypothesised, and an independent opportunity for testing it would be of interest.

%--------------------
\section{The barycentric motion of exoplanet host stars}
%--------------------

Multiple exoplanet systems, those with two or more known orbiting planets, provide an independent possibility of testing the hypothesis that the Sun's barycentric motion is somehow linked to solar activity. To do this, specific features of their host star barycentric motions (as determined through Doppler measurements in the radial direction) can be compared with their own stellar activity indicators. This study is concerned with the former aspect.

Of nearly 500~known exoplanets, 41~multiple systems discovered by radial velocity measurements are currently known. Of these, 30~systems have two known planets, while 11~systems have three or more. This sample of Doppler-detected multiple systems is complemented by the two multiple systems discovered from photometric transits, CoRoT--7 and HAT--P--13.

To determine the host star barycentric motions, we have used the orbital elements for the multiple systems compiled at {\tt www.exoplanets.org}. The characteristics of the Doppler-detected systems according to stellar mass, planetary mass, orbital separation and eccentricity, are shown schematically (but to scale) in Figure~\ref{fig:multiples}. The eccentricity, $e$, semi-major axis, $a$, and argument of pericentre, $\omega$, define the individual orbit geometries. The time of periastron passage, $t_{\rm p}$, and orbital period, $P$, additionally determine the position of the planet along its orbit. 

We assume lower limits for the planetary masses, i.e.\ $M_{\rm p}\sin i$ with $\sin i=1$ (giving lower limits to the stellar displacements), and further that all planets in a given system are coplanar. We ignore dynamical planetary interactions known to be significant in some cases, and quadrupole or relativistic precession, as irrelevant for the qualitative orbital properties over relatively short time intervals. For the Sun, only the contributions of Jupiter, Saturn, Uranus and Neptune are considered. Their eccentricities are included, although they have little additional effect.

%--------------------
\subsection{Orbital angular momentum}
%--------------------

In the orbit plane, the contribution of the star's displacement with respect to the system barycentre $\bf r$, its velocity $\bf v$, and its acceleration~$\dot{\bf v}$, are calculated for each planet $i$, under the assumption $M_i\ll M_\star$, as
\begin{eqnarray}
{\bf r}_i &=& \frac{M_i \, a_i}{M_{\rm tot}}
		 \left(\! \begin{array}{c}
			\cos\omega(e-\cos E)+\sqrt{1-e^2}\sin\omega\sin E\\
			\sin\omega(e-\cos E)-\sqrt{1-e^2}\cos\omega\sin E\\
			0
		\end{array} \! \right) \\
{\bf v}_i &=& \frac{2\pi M_i \, a_i}{M_{\rm tot} P(1-e\cos E)} 
	  \left(\! \begin{array}{c}
			\cos\omega\sin E+\sqrt{1-e^2}\sin\omega\cos E\\
			\sin\omega\sin E-\sqrt{1-e^2}\cos\omega\cos E\\
			0
		\end{array} \! \right) \\
\dot{\bf v}_i &=& -\frac{4\pi^2 M_i \, a_i}{M_{\rm tot} \, P^2(1-e\cos E)^3} \nonumber \\
		&& \qquad \left(\! \begin{array}{c}
			\cos\omega(e-\cos E)+\sqrt{1-e^2}\sin\omega\sin E\\
			\sin\omega(e-\cos E)-\sqrt{1-e^2}\cos\omega\sin E\\
			0
		\end{array} \!\right) \ ,
\end{eqnarray}
where $M_i$ are the individual planet masses, $M_{\rm tot}=M_\star+\sum M_i$, and $a_i$ are the semi-major axes of the planet's orbit around the star (not the barycentre).
In these equations, the eccentric anomaly~$E$ at time~$t$ is derived iteratively from Kepler's equation,
\begin{equation}
E-e\sin E = \frac{2\pi}{P}(t-t_{\rm p}) \ .
\end{equation}
The total angular momentum~$\bf L$ of the star with respect to the system barycentre, and its time derivative (identically zero for a single planet system), are then given by
\begin{eqnarray}
{\bf L} 					&=& M_\star \; {\bf r}\times {\bf v} \ ,\\
\frac{{\rm d}{\bf L}}{{\rm d}t} 	&=& M_\star \; {\bf r} \times \dot{\bf  v} \ ,
\end{eqnarray}
where ${\bf r}$, ${\bf v}$, and $\dot{\bf v}$ are the sums over the individual contributions, from which we derive the components perpendicular to the common orbital plane
\begin{eqnarray}
L_z 						&=& M_\star (r_x v_y-r_y v_x) \ , \\
\frac{{\rm d}L_z}{{\rm d}t}		&=& M_\star (r_x \dot v_y-r_y \dot v_x) \ .
\end{eqnarray}

In the accompanying plots, the $xy$-projection corresponds to the orbital plane, and the $x$-axis points from the barycentre towards the common ascending node. In the case of the solar system, the $xy$-plane essentially corresponds to the ecliptic, and the $x$-axis points from the barycentre towards the equinox.

%-----------------------------------
\begin{figure*}[t]
\centering
\includegraphics[width=0.99\linewidth]{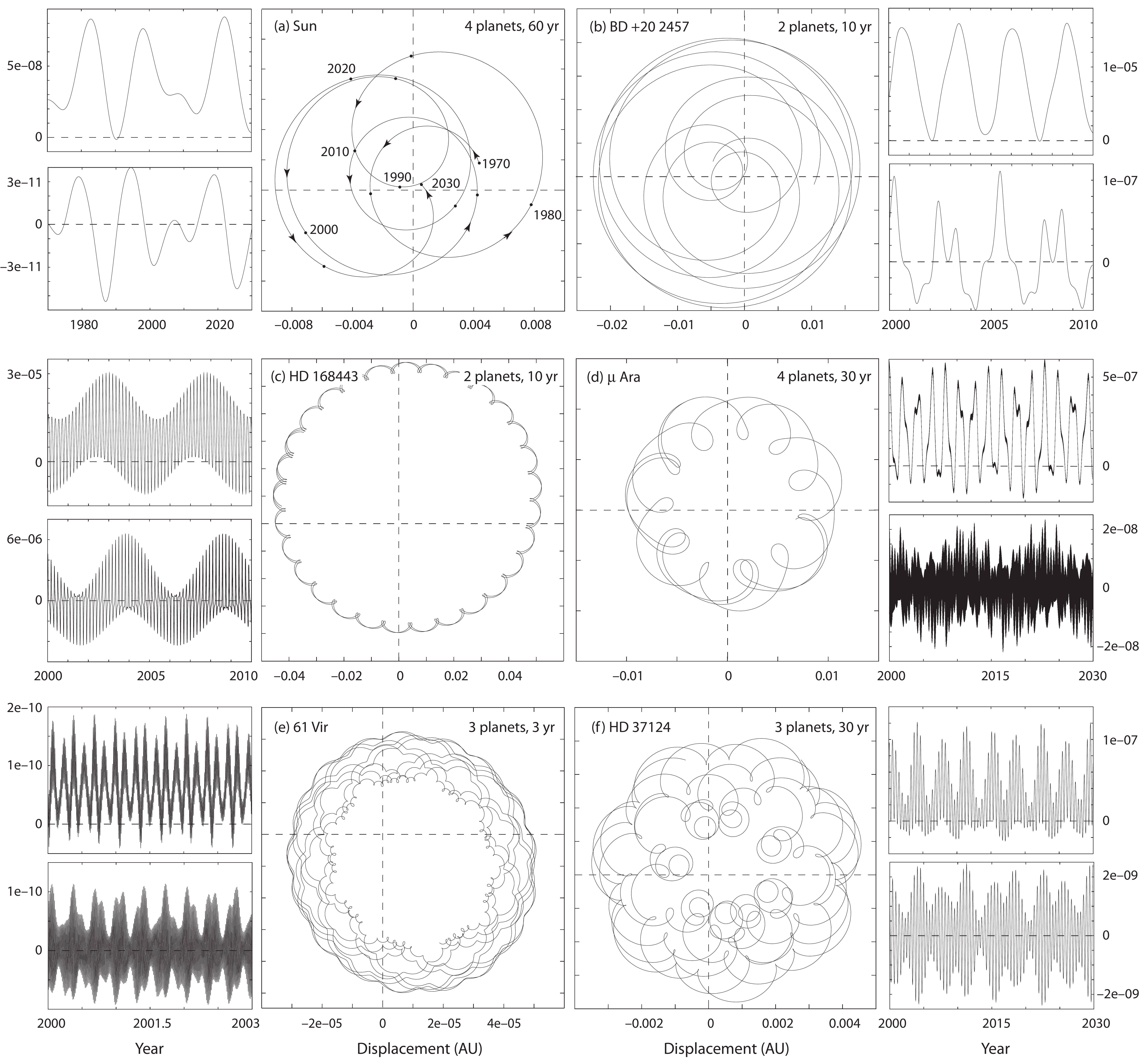} 
\caption{Barycentric motion of the host star for a selection of representative multiple exoplanet systems. Main plots (central two columns) show the orbit over the indicated time interval in a reference frame with the system barycentre at the origin, with abscissae and ordinates in~AU (the solar diameter is $R_\Sun=6.96\times10^8$\,m, or 0.00465\,AU). To the outer side of each orbital sequence, plots show the orbital angular momentum, $L_z$ (upper), and ${\rm d}L_z/{\rm d}t$ (lower) for the same time interval, in the units as given in Table~\ref{tab:systems}. For the systems shown, star masses lie in the range $0.85-1.15M_\Sun$ except for BD~+20\,2457 which is $\sim$2.8$M_\Sun$. Orbital parameters were taken as follows:
the Sun: \citet{2005seidelmann}; % in practice, from http://ssd.jpl.nasa.gov/?planet\_pos\#formulae
BD~+20\,2457: \citet{2009ApJ...707..768N};
HD~168443: \citet{2009ApJ...693.1084W};
$\mu$~Ara: \citet{2007A&A...462..769P};
61~Vir: \citet{2010ApJ...708.1366V};
and HD~37124: \citet{2005ApJ...632..638V}.
\label{fig:orbits}
}
\end{figure*}
%-----------------------------------

%-----------------------------------
\begin{figure*}[t]
\centering
\includegraphics[width=0.95\linewidth]{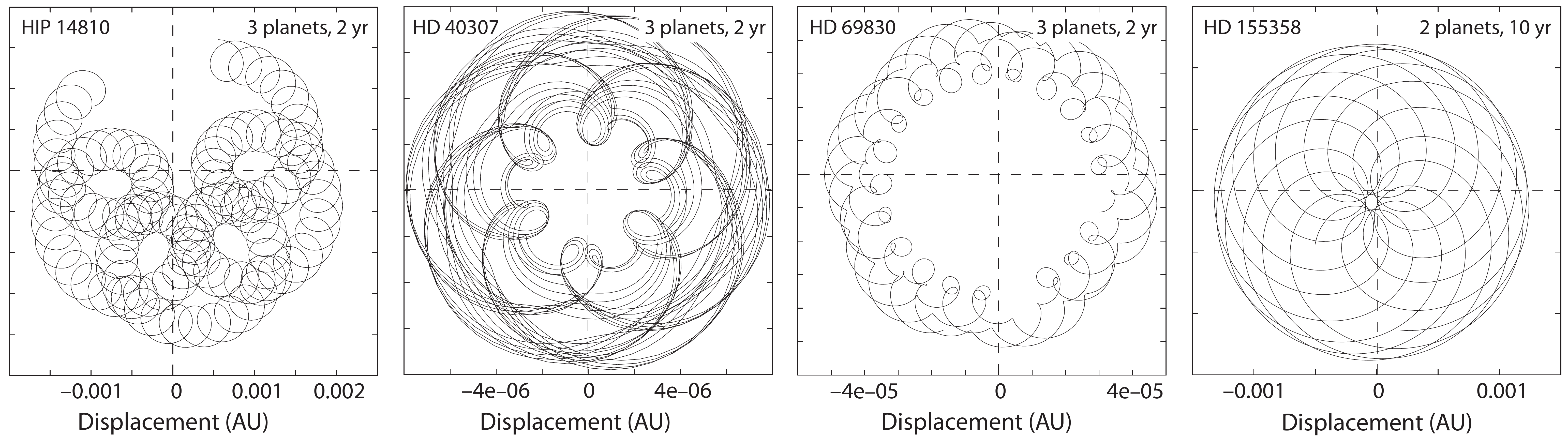} 
\caption{As Figure~\ref{fig:orbits}, showing further examples of barycentric orbits for
HIP~14810,
HD~40307,
HD~69830,
and HD~155358.
\label{fig:orbits-other}
}
\end{figure*}
%-----------------------------------

%--------------------
\section{Results}
%--------------------

Table~\ref{tab:systems} summarises the results from the dynamical analyses, ordered by increasing $\vert\,{\rm d}L_z/{\rm d}t\,\vert_{\rm max}$ determined over 100~years from 2000--2100. All but two of the stars have $\vert\,{\rm d}L_z/{\rm d}t\,\vert_{\rm max}$ exceeding that of the Sun. HD~168443 has the largest ${\rm d}L_z/{\rm d}t$, more than five orders of magnitude larger than that of the Sun.

The majority of stars, 26 out of~41, also experience periods of negative orbital angular momentum, $L_z$. HD~168443 is furthermore the system with the smallest (most negative) value, some four orders of magnitude larger than the negative excursion of the Sun (not present in the time interval of Table~\ref{tab:systems}). Stars with the next most negative extrema are HD~74156 and HD~217107.

%--------------------
\subsection{Orbital motions}
%--------------------

Figures~\ref{fig:orbits}--\ref{fig:orbits-other} show a selection of the resulting host star barycentric paths, including that of the Sun. The behaviour can be compared with the architecture of the systems shown in Figure~\ref{fig:multiples}.

The path of the Sun (Figure~\ref{fig:orbits}a) replicates that of earlier work \citep[e.g.][]{1965AJ.....70..193J, 2000SoPh..191..201J}, with maximum displacements exceeding $1R_\Sun$. Our determinations of $L_z$ and ${\rm d}L_z/{\rm d}t$ also closely match those of \citet[][their Figure~2b]{1965AJ.....70..193J} and \citet[][their Figure~1]{2005MNRAS.362.1311J}. An excursion to negative $L_z$ occurs around 1990.

A number of systems, including those dominated by multiple planets with $a\le 0.2-0.3$\,AU, result in only small perturbations from a predominantly elliptical stellar path, even for some triple or quadruple planetary systems. Examples (not illustrated) are 
GJ~876, 
BD~$-$08\,2823, 
HD~11964,
HD~38529,
HD~47186,
HD~128311,
HD~187123,
HD~190360,
HD~215497
and HD~217107.
The five-planet system 55~Cnc is qualitatively similar, displaying essentially circular motion dominated by the most massive outer planet ($4M_\Jupiter$ at 6\,AU), but with low-amplitude high-frequency perturbations due to the second most massive planet ($0.8M_\Jupiter$ at 0.1\,AU). 

Slightly more complex motion is seen in 
47~UMa, 
HD~9446,
HD~45364,
HD~60532,
HD~73526,
HD~82943,
HD~202206, 
as well as in the transiting multiple
CoRoT--7.

BD~+20\,2457 (Figure~\ref{fig:orbits}b) is the most massive star in the present sample at $\sim$2.8$M_\Sun$. With two massive planets ($23M_\Jupiter$ at 1.5\,AU and $13M_\Jupiter$ at 2\,AU), it displays a barycentric motion qualitatively most similar to that of the Sun. 

HD~168443 (Figure~\ref{fig:orbits}c) has two massive planets at 0.3--3\,AU ($8-18M_\Jupiter$), with a high eccentricity $e\sim0.5$ of the inner planet. The star shows the most extreme values of negative $L_z$ and largest ${\rm d}L_z/{\rm d}t$, despite its relatively simple orbit. These excursions are regular, with a periodicity  of $\sim$58\,d given by the period of the innermost planet.
Similar motion is seen in
HD~74156,
HD~147018,
$\upsilon$~And,
and in the wide-transiting HAT--P--13.

For $\mu$~Ara (Figure~\ref{fig:orbits}d), a periodic looping motion is dominated by the two most massive outer planets ($1.9M_\Jupiter$ at 5.3\,AU and $1.7M_\Jupiter$ at 1.5\,AU).  The third most massive planet ($0.5M_\Jupiter$ at 0.9\,AU) imposes more abrupt changes in orbital angular momentum on the underlying two-planet induced motion.
Similar effects are seen in 
HD~12661,
HD~108874,
HD~134987,
HD~169830,
HD~181433
and HD~183263.
In HD~181433, the eccentricities of the two most massive outer planets result in a host star motion systematically displaced from the system barycentre.

For 61~Vir (Figure~\ref{fig:orbits}e), the three planets are of low mass ($0.01-0.07M_\Jupiter$), and lie at small separation (0.05--0.5\,AU). While the motion is complex, the barycentric displacements and angular momentum changes are all small. GJ~581 is similar.

HD~37124 (Figure~\ref{fig:orbits}f) is a triple system with almost identical planetary masses in the range $0.6-0.7M_\Jupiter$. A simple looping motion like that of $\mu$~Ara results from the two most massive planets at 0.5 and 3.2\,AU. But this is transformed into a complex spiralling pattern in the presence of the third planet at 1.7\,AU, arising from the various planetary orbital periods of 154, 844, and 2300\,d.
Somewhat similarly complex but low-amplitude structure is seen in a number of other systems, including 
HIP~14810,
HD~40307,
HD~69830,
and HD~155358 (Figure~\ref{fig:orbits-other}).

On a historical note, a curve similar to the symmetric part of Figure~\ref{fig:orbits}d, illustrating the orbit of Mars viewed from Earth, appeared in Kepler's 1609 {\it Astronomia Nova}. These families of curves have been named `planet mandalas' by N.~Michelsen \& M.~Pottinger, after the Sanskrit for circle, and appear as such in the Wolfram Demonstrations Project, http://demonstrations.wolfram.com/SolarSystemMandalas/.
 
%--------------------
\section{Correlation with stellar activity}
%--------------------

A detailed investigation of any correlation between these motions and stellar activity lies beyond the scope of this paper. But we note that observable diagnostics of stellar activity are available in principle, and are becoming so in practice. Amongst these are radial velocity `jitter'-type measurements probing stellar atmospheric inhomogeneities such as spots and plages modulated by stellar rotation \citep[e.g.][]{1997ApJ...485..319S, 1998ApJ...498L.153S, 2006MNRAS.372..163J, 2007A&A...473..983D, 2009AIPC.1094..152S}; chromospheric activity measured through the Ca\,{\footnotesize II} H~and~K lines \cite[e.g.][]{2001MNRAS.325...55S, 2004ApJS..152..261W}, sometimes attributable to tidal or magnetic interactions \citep{2000ApJ...533L.151C}, with some monitoring programmes extending back to 1966 \citep{1995ApJ...438..269B}; star spots observed photometrically during transits \citep[e.g.][]{2008ApJ...683L.179S, 2009ApJ...701..756D}; and changes in photospheric radius due to varying magnetic activity \citep{2009NewA...14..363L}.

For the F7 star $\tau$~Boo, a $4.1M_\Jupiter$ planet orbits in a 3.31\,d period with $a_{\rm p}=0.048$\,AU \citep{2006ApJ...646..505B}. The star has a relatively shallow convective envelope of about $0.5M_\Jupiter$ \citep{2005ApJ...622.1075S, 2008ApJ...676..628S}. It has a weak magnetic field of $1-3\times10^{-4}$\,T \citep{2007MNRAS.374L..42C}, which underwent a global magnetic polarity reversal between 2006--07 \citep{2008MNRAS.385.1179D}, and again between 2007--08 \citep{2009MNRAS.398.1383F}, switching between predominantly poloidal and toroidal forms. Observations suggest a magnetic cycle of $\sim$2\,yr, much shorter than the Sun's 22\,yr. 

\citet{2009MNRAS.398.1383F} hypothesise that this is governed by the close-orbiting planet, through tidal synchronisation of the outer convective envelope enhancing the shear at the tachocline. 

From our study \object{HD~168443}, with the innermost planet at $a_{\rm p}\sim0.3$\,AU, offers particularly interesting prospects for investigation as to whether much more distant planets can affect stellar activity through the star's reflex motion. Present physical understanding says that they cannot. But its ${\rm d}L_z/{\rm d}t$, with a periodicity of 58\,d, exceeds by more than five orders of magnitude that of the Sun. If orbital angular momentum variations play a role, its effects should be visible in this system.

\object{HD~74156} displays the next most negative orbital angular momentum intervals. Qualitatively similar to HD~168443 (Figure~\ref{fig:orbits}c) these negative excursions extend over 27 or so 51-d periods (3.8\,yr), and are then absent for the subsequent 20 or so periods (2.8\,yr). Its large mass, $1.24M_\Sun$, further implies a shallow convective envelope, perhaps with greater susceptibility to shear effects. If retrograde angular momentum plays a role, these specific modulations will offer a defining diagnostic.

Such periodicities observed in their stellar activity would support the unorthodox picture that the barycentric motion of the Sun somehow affects long-term activity cycles. Otherwise, it would imply that any link between the two is spurious.

Much larger orbital angular momentum is experienced for stars in wide binary orbits, including exoplanets in circumprimary (S-type) orbits such as that around 16~Cyg~B, where the stellar binary has $P_{\rm orb}\sim18\,000$\,yr. Heuristically, it is difficult to imagine that $L_z$ alone can have any relevance.

%--------------------
\section{Conclusions}
%--------------------

Our study demonstrates that a variety of complex barycentric motions exists for exoplanet host stars. Behaviour cited as being correlated with the Sun's activity, for example intervals of more disordered motion, large changes in orbital angular momentum, and intervals of negative orbital angular momentum, are common -- but more extreme -- in exoplanet systems. 

Accompanied by detailed studies of the associated stellar activity, these systems offer an independent opportunity to corroborate the hypothesised link between the Sun's barycentric motion and the many manifestations of solar activity. In particular, they offer the possibility of independently testing any theories of spin--orbit coupling which are advanced in the case of the Sun.

\begin{acknowledgements}
Michael Perryman acknowledges support from the University of Heidelberg and the Max Planck Institute for Astronomy, Heidelberg, in the form of a Distinguished Visitor award. Tim Schulze--Hartung gratefully acknowledges a grant from the Max Planck Society. We thank Lennart Lindegren for valued comments on an earlier draft, and the referee, Fran\c cois Mignard, for important clarifications in the approximations used.
\end{acknowledgements}

\end{document}